\documentclass[12pt]{article} % single column, double spaced

\usepackage{ol2}
\usepackage{hyperref}
\usepackage{amsmath}

\begin{document}

\twocolumn[ %% activate for two-column option

\title{\textit{} Asymptotic estimations of power thresholds and anti-Stokes frequency of laser induced thermal scattering in microresonators}

\author{M.V.Jouravlev}

\address{Raymond and Beverly Sackler Faculty of Exact Sciences, School of Chemistry,
Tel-Aviv University, \\ Ramat-Aviv, 69978, Israel}

% Do not use \email or \homepage here. E-mail and URL can be given just before references.

\begin{abstract} A derivation of the asymptotic expressions for the threshold intensity
of laser induced thermal scattering in silica microresonator when
illuminated with a plane wave is present. The calculation of
anti-Stokes thermal combination frequencies are made for the
spherical high Q-factor microresonators. The three modes regime of
nonlinear interaction is considered. One pump-driven, two signal
modes, and one mode of temperature relaxation are taken into
account, satisfying morphology-dependent input and output
resonances. There are low power thresholds for laser induced thermal
scattering at morphology dependent resonances.
\end{abstract}

\ocis{140.3410, 140.6810}

 ] %% activate for two-column option

\noindent

Stimulated thermal Rayleigh and thermal Mandelshtam-Brillouin
scattering from bulk liquids have recently been
reported\cite{Herm,Mack,Besp,Batra}. The observations have been
treated theoretically in terms of entropy and localized thermal
fluctuations \cite{Blom}. Rapid thermalization of the molecules
which absorb at the laser pump, and the subsequent generation of
large density or temperature fluctuation, leads to the enhancement
of the thermal density fluctuations and increase the anti-Stokes
frequency shift in bulk. There are numerous thermal nonlinear
effects in silica microresonators due to the extremely high Q-factor
of electromagnetic modes coinciding with "Morphology Dependent
Resonance" conditions (MDR)\cite{Chang,Gor3,Gor4}. At the specific
threshold intensity of the laser pump and at the resonant size
parameter of the cavity, the surface layer of the cavity
significantly enhances the internal field inside the cavity at the
incident resonant wavelength of laser pump, and further more
efficiently provides optical thermal feedback from the internally
generated wave through optical bistability and instability
effects\cite{Vah2,Vah3,Gor1,Gor2}. The time dependent shift of the
eigenfrequencies of the "Whispering Gallery Modes" (WGM's) can be
caused by intensity-dependent mechanisms, such as the change in the
refractive index which is dependent on laser light intensity and
caused by an absorption-induced temperature change. The time
dependent shift of the eigenmode frequencies is due to heating by
laser power absorption, creating periodic thermal oscillations
(oscillation regime)in spherical high Q-factor silica
resonators\cite{Gor1,Gor2} and aperiodical oscillation (bistability
regime) of scattering amplitudes, which have been experimentally
investigated in microtoroid resonators\cite{Vah2,Vah3}. These
experimental observations of the threshold oscillations and
  gain enhancements are associated with thermal nonlinear
  MDR and nonlinear interaction of WGM's in microspheres. The purpose
of this communication is to report the calculation of thresholds of
a new stimulated effect, which is believed to be attributable to
laser induced stimulated thermal scattering in microresonators. The
effects of the laser induced thermal scattering (LITS) in
microresonators at MDR conditions has not been discussed in the
literature. This article reports new knowledge relating to LITS,
applicable to microresonators. The dielectric permeability of the
resonator substance depends on the temperature of the resonator and
energy in the volume of interacting modes which absorbed the laser
light power. The dielectric permeability of the microresonator
determines the relative shifts of the eigenmodes of the
microresenator. A change in the dielectric permeability leads to a
shift in the eigenmodes excited by the laser pump in the
microresenator. Furthermore, independently, due to the heating of
the resonator by the laser pump in input MDR conditions, the
eigenfrequencies of the microresenator also have a periodic thermal
shift. The maximum absorbtion of the laser power and heating occurs
at MDR frequencies and a positive shift of the eigenfrequencies
occurs when heat is released from the surface of the resonator.
Alternatively, when heat is generated at the surface of a resonator,
the eigenfrequencies of the resonator take on a negative shift. The
thermal coupling of the electromagnetic modes in a microresonator
provides the appearance in the scattering spectra of additional
anti-Stokes thermal frequency shifts. One would expect a lower
threshold intensity of LITS to be caused by thermal mode
 overlapping and MDR conditions simultaneous with Raman
 lasing. The lowering of the threshold of the Raman laser emission in silica
resonators (with radius $R = 40$ $\mu m$) has been
reported\cite{Vah1}. Early theoretical explanations and the
estimations of the low power thresholds for Raman lasing in
microspheres have also been reported \cite{Kur2,Braun}.
 Three modes regime of interaction were considered. The first mode is
the thermal mode, the second is the signal and the third is the pump
mode. Maxwell's electromagnetic equations and thermal equations for
one thermal mode were solved by applying the methods of slow varied
amplitudes for the system of ordinary differential oscillation
equations\cite{Bel1}. If we take into account the partial wave
amplitudes of the thermal oscillations, and compute the scattering
power of a resonator,(introducing Q-factor as the ratio of the field
energy inside the mode to incident power, and multiplying by the
leakage rate for power threshold) then we set\cite{Bel2}:

\begin{align}
P_{th}=\frac{\omega_{p}^{2}}{2\pi Q_{i}Q_{j}K_{ijk}H_{ijk}}
\frac{k}{\rho C_{p}}\left(\frac{\mu_{i}}{R}\right)^{2}
\end{align}

Here: $K_{ijk}$ and  $H_{ijk}$ are the thermal and electromagnetic
mode overlapping coefficients estimated below, $i$ corresponds to
the thermal mode, $j$ and $k$ correspond to the electromagnetic
eigenmodes; $\omega_{p}$ is the pump frequency, $Q_{i}$ and $Q_{j }$
are the Q-factors of $i$ and $j$ eigenmodes, $\mu_{i}$ is the
eigenvalue of thermal conductivity equation\cite{Bel2}, $\rho$ is
the density of the resonator, $k$ is the thermal conductivity,
$C_{p}$ is the specific heat capacity, $R$ is the radius of
resonator. The threshold power is obtained at MDR conditions for the
optimal tuning conditions: $2(\tau_{t}+\tau_{e})
=1+\omega_{f}^{2}/\omega_{p}^{2}$, where
$\tau_{t}=k(\mu_{i}/R)^{2}/\rho C_{p}\omega_{p}$ and
$\tau_{e}=1/2Q_{i}$ are the relaxation times of thermal and
electrical modes with $\omega_{f}$. Electromagnetic decay rate of
the effective three-mode interaction is $\tau_{e}$ and the thermal
relaxation rate is $\tau_{t}$. The frequency $\omega_{f}$ can be
calculated by the transcendental eigenvalue equations of Mie's
theory using the resonant size parameter at MDR. For the optimal
tuning conditions and the lowering of the LITS threshold by MDR
conditions, the thermal anti-Stokes frequency is\cite{Bel2}:
\begin{align}
\Omega_{T}=D\left(\frac{\mu_{i}}{R}\right)^{2}
\end{align}
Here: $D =k/\rho C_{p}$. Although two high Q-factor circuits are
present in the microresonator, the optimal detuning during
three-mode LITS can be small if the frequency interval between the
resonator modes is approximately equal to $\Omega_{T}$. WGM's
efficiently provide the thermal feedback for internally excited
modes at the specific frequency matching:
$\omega_{p}=\omega_{s}\pm\Omega_{T}$.
 Equations (1),(2) provide the information concerning the absorbed substance
  of a resonator as well as the yield of the multi-exponential thermal
damping of any stimulated processes at times comparable with a
thermal oscillation in Eq.(2). The mode overlapping coefficients
$K_{ijk}$ and $H_{ijk}$ have a complicated structure, and consist of
the high oscillating eigenfunctions\cite{Bel2}.
 The mode overlapping coefficients have the following form \cite{Bel1}: $H \simeq
\omega_{f}^{2}a_{\varepsilon}T$, $T^{2}\simeq(\rho C_{p}V)^{-1}$ and
$K \simeq \sigma T$ where: $\sigma=\epsilon\omega_{f}/4\pi Q_{i}$,
$\epsilon$ is the dielectric permeability, $a_{\varepsilon}$ is the
temperature coefficient of dielectric permeability, $V$ is the
effective volume of interaction of WGM's\cite{Gor3}. The
calculations were made by using the known values of material
parameters valid for the experiments: density of fused silica
$\rho=2.21$ $[g/cm^3]$, thermal conductivity $k=1.4\cdot10^{-2}$
$[W/cmK]$, specific heat capacity $C_{p}=0.67$ $[Ws/gK]$,
$a_{\varepsilon}=1.45\cdot10^{-5}$ $[K^{-1}]$, $D=9.5\cdot10^{-3}$
$[cm^{2}/s]$, refraction index- $n=1.46$, $Q_{i}=10^8$\cite{Gor4}.
 An asymptotic estimate of the mode overlapping integral is valid for a high
index of partial wave amplitudes of WGM's $n\simeq\ x$ (here $x$ is
the resonant size parameter) for n surface modes obtained from
Cauchy-Bunyakowski-Schwarz inequality\cite{Bel1}.
 The mode overlapping coefficients $K$ and $H$ were written under the following conditions: (1) small
losses of WGM's, which are appropriate for the high Q-factor of the
resonator modes; (2) singling out two interacting modes within the
homogeneous thermal mode volume frequency interval. One is the pump
mode and the other is the resonant signal (anti-Stokes) mode,
namely, both input and output resonance conditions\cite{Chang}. The
input resonance condition is satisfied for a broadband thermal
detuning of the input and output modes, which spans several high-Q
MDR's, whereas the output resonance condition is always satisfied,
since the bandwidth of LITS spans at least several high-Q MDR's.
Substituting the asymptotic estimation for the mode overlapping
coefficients $K$ and $H$ in Eq.(1) for the threshold power of LITH
yields:
\begin{align}
P_{th}=\frac{2kV}{\varepsilon
a_{\varepsilon}Q_{i}}\left(\frac{\mu_{j}}{R}\right)^{2}
\end{align}
The threshold of LITS is then determined by two circumstances: the
effective resonant heating absorption and the overlapping mode
coefficients\cite{Bel2}. In Fig.1 the dependence of the threshold
power on the fused silica spherical resonator radii is shown. The
interacting nondegenerate WGM's $"TE"$,$(TE_{n},n\simeq\ x)$,($x$-
resonant size parameter) are coupled to thermal modes by thermal
nonlinearity at wavelength of $840$ $nm$. The computed threshold
input intensities in a silica nonlinear sphere vary from $2$ to $23$
$\mu W$ for radii of $40$ to $110$ $\mu m$, the Q-factor of
 a $"TE"$ resonant mode is $Q_{i}=10^{8}$\cite{Gor4}.
\begin{figure}[htb]
\centerline{\includegraphics[width=8cm]{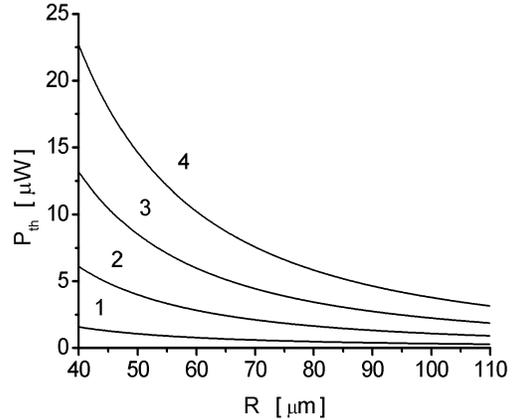}} \caption{The
threshold power of STS for fused silica microsphere. Radius
 and thermal mode dependence: 1. $ TE-T_{1}^{1}$ modes, 2.
$TE-T_{3}^{1}$ modes, 3. $TE-T_{5}^{1}$ modes, 4. $TE-T_{7}^{1}$
modes. The laser pump is mode-locked $Ti:Al_{2}O_{3}$ laser
operating at $840$ $nm$ with the pulse repetition frequency $82$
$MHz$ and the pulse duration $1$ ps.}
\end{figure}
\begin{figure}[htb]
\centerline{\includegraphics[width=8cm]{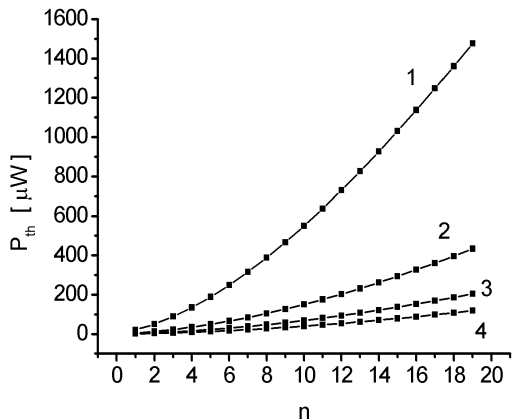}} \caption{ The
threshold power of LITS of $T_{n}$ modes with radius of sphere $R =
40$ $\mu m$ 1. $TE-T_{n}^{1}$, 2. $TE-T_{n}^{2}$, 3. $TE-T_{n}^{3}$,
4. $TE-T_{n}^{4}$. The pump is at $1.55$ $\mu m$ of external CW-
diode laser with $300$ $kHz$ line-width\cite{Vah1}.}
\end{figure}
As illustrated in Fig.2, the threshold incident pump power $P_{th}$
of LITS depends on the thermal $T_{n}^{1}$ mode order. More
quantitatively  the threshold pump power for LITS Eq.(3) is found to
be less $50$ $\mu W$, $1\leq n\leq8$, for the radius of $40$ $\mu
m$, well below the threshold of Raman lasing $86$ $\mu W$
\cite{Vah1} and stimulated Brillouin scattering in glass sphere,
which has the order of $160$ $W$ \cite{Cantrell}. To be specific,
the threshold power of LITS is less then the threshold of stimulated
Raman lasing and stimulated Brillouin scattering in silica
microspheres, thus the thermal interaction of the eigenmodes can be
revealed by the spectra of the Raman lasing with thermal instability
or bistability in microresonators.
\begin{figure}[htb]
\centerline{\includegraphics[width=8cm]{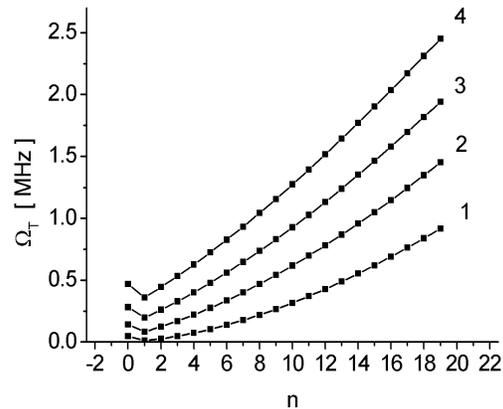}} \caption{The
Anti-Stokes combination frequency of LITS from fused silica
microsphere irradiated by CW-diode laser pump at $1.55$ $\mu
m$\cite{Vah1},$R=40$ $\mu m$ 1. $TE-T_{n}^{4}$,
 2. $TE-T_{n}^{3}$, 3. $TE-T_{n}^{2}$, 4.
 $TE-T_{n}^{1}$.}
\end{figure}
As illustrated in Fig.3, the thermal anti-Stokes combination
frequencies are dependent on the number of the interacting modes.
The thermal anti-Stokes combinational frequency of LITS can be
varied over a wide range of frequencies from Rayleigh's frequency of
shift via thermal Rayleigh scattering \cite{Batra} to the striction
combination frequency of thermal Mandelshtam-Brillouin
 scattering\cite{Mack,Blom}. Such a wide range of
anti-Stokes frequency can be provided by systems such as a spherical
resonator. The anti-Stokes frequencies of LITS can occur in spectra
of Mandelshtam-Brillouin, Raman scattering and linear Mie's
scattering. The frequency can be varied over the wide range and
depends on the material parameters of the resonator as well as both
the effective volume of WGM and the surface mode overlapping. In
order to have the description of Raman scattering and lasing in
agreement with the experiments it would be necessary to take into
account the thermal modes in the set of electromagnetic equations.
The difference in the threshold power of interacting thermal modes
enables the separation of these nonlinear processes from each other.
The thermal oscillations and the thermal spatial overlapping of
electromagnetic modes are important processes for Raman scattering
and lasing in microresonators.

\end{document}